\documentclass[prl,twocolumn,preprintnumbers,superscriptaddress,amsmath,amssymb]{revtex4-2}%

\usepackage{hyperref}
\usepackage{epsf,epsfig}
\usepackage{amssymb,amsmath,amsfonts}
\usepackage{color}
\usepackage{braket}
\usepackage{subcaption}
\def\pb[#1,#2]{\{#1, #2\}}
\def\deb[#1,#2]{[#1,#2]_{\text{D.B.}}}

\def\sl{\slashed}

\def\Or[#1]{{\text{O}}\left({#1}\right)}
\def\dotl[#1,#2]{\left\langle #1,\, #2 \right\rangle}
\def\dotlb[#1,#2]{\left\langle #1,\, #2 \right\rangle}
\def\dotlm[#1,#2]{\left[ #1,\, #2 \right]}
\def\dotp[#1,#2]{(\vect{#1} \cdot\vect{#2})}
\def\aff[#1,#2]{\hat{#1}(#2)}

\def\n4sym{{\cal N}=4 SYM}
\def\>{\rangle}
\def\<{\langle}
\def\weight[#1,#2,#3]{\{(#1),#2,#3\}}
\def\ads[#1]{$\text{AdS}_{#1}$}

\def\tarelr[#1]{\widetilde{a}^{\text{rel}}_{R#1}}
\def\Oright[#1]{{\cal O}_{R#1}}
\def\Oleft[#1]{{\cal O}_{L#1}}
\def\aleft[#1]{a_{L#1}}
\def\arelr[#1]{a^{\text{rel}}_{R#1}}

\def\Tr{{\rm {Tr}}}
\def\hkll{{\Phi}}

\newcommand{\be}{\begin{equation}}
\newcommand{\ee}{\end{equation}}
\newcommand{\beq}{\begin{equation}}
\newcommand{\eeq}{\end{equation}}
\newcommand{\ben}{\begin{displaymath}}
\newcommand{\een}{\end{displaymath}}
\newcommand{\beqa}{\begin{eqnarray}}
\newcommand{\eeqa}{\end{eqnarray}}
\newcommand{\bea}{\begin{eqnarray}}
\newcommand{\eea}{\end{eqnarray}}
\newcommand{\bean}{\begin{eqnarray*}}
\newcommand{\eean}{\end{eqnarray*}}
\newcommand{\ba}{\begin{array}}
\newcommand{\ea}{\end{array}}
\newcommand{\bi}{\begin{itemize}}
\newcommand{\ei}{\end{itemize}}

\newcommand{\bs}{\begin{split}}
\def\sess\end{split}

\newcommand{\vect}[1]{{#1}}

\def\rsz{\ket{\Psi_{0}}}
\def\lsz{\bra{ \Psi_0}}

\def\Tr{{\rm Tr}}

\def\drop{\widehat{\Phi}}

\begin{document}

\title{State-dressed local operators in the AdS/CFT correspondence}

\preprint{CERN-TH-2022-145}

\author{Eyoab Bahiru}
\email{ebahiru@sissa.it}
\affiliation{SISSA, International School for Advanced Studies, via Bonomea 265, 34136 Trieste, Italy}
\affiliation{INFN, Sezione di Trieste, via Valerio 2, 34127 Trieste, Italy}
\affiliation{International Centre for Theoretical Physics
Strada Costiera 11, Trieste 34151 Italy}
\author{Alexandre Belin}
\email{alexandre.belin@unige.ch}
\affiliation{Department of Theoretical Physics, University of Geneva, 24 quai Ernest-Ansermet, 1211 Geneva 4, Switzerland}
\affiliation{Theoretical Physics Department, CERN, CH-1211 Geneva 23, Switzerland}
\author{Kyriakos Papadodimas}
\email{ kyriakos.papadodimas@cern.ch}
\affiliation{Theoretical Physics Department, CERN, CH-1211 Geneva 23, Switzerland}
\author{G\'abor S\'arosi}
\email{gabor.sarosi@cern.ch}
\affiliation{Theoretical Physics Department, CERN, CH-1211 Geneva 23, Switzerland}
 \author{Niloofar Vardian}
\email{nvardian@sissa.it}
\affiliation{SISSA, International School for Advanced Studies, via Bonomea 265, 34136 Trieste, Italy}
\affiliation{INFN, Sezione di Trieste, via Valerio 2, 34127 Trieste, Italy}

\begin{abstract}
\noindent
We examine aspects of locality in perturbative quantum gravity and how information can be localized in subregions. In the framework of AdS/CFT, we consider the algebra of single-trace operators defined in a short time band. We conjecture that if the state has large energy variance, then this algebra will have a commutant in the $1/N$ expansion.  We provide evidence for this by identifying operators that commute with the conformal field theory Hamiltonian to all orders in $1/N$, thus resolving an apparent tension with the gravitational Gauss law. The bulk interpretation is that these operators are gravitationally dressed with respect to features of the state rather than the boundary. 
We comment on observables in certain black hole microstates and the gravitational dressing in the island proposal.
\end{abstract}
\maketitle

\section{I. Introduction}

Locality in nongravitational quantum field theory (QFT( is well understood. It can be expressed by the axiom of microcausality \cite{Streater:1989vi} or more broadly by the structure of ``net of algebras'' \cite{Haag:1963dh, Haag:1992hx}. While it is not straightforward to factorize the Hilbert space into subsystems corresponding to spacelike separated regions, due to UV-divergent entanglement and the type III$_1$ nature of local algebras \cite{araki},  there is a way of thinking about subsystems in terms of the split property \cite{Buchholz:1973bk}. In particular, on any given time slice quantum information can, in principle, be strictly localized in finite spatial regions.

Classical general relativity also respects the principle of locality. While the constraints in the Hamiltonian formulation impose nontrivial conditions on the initial value problem allowing some properties of the state, like the total mass, to be read off from infinity,  it is still possible to localize information in subregions of space; see, for example, \cite{Giddings:2019hjc} for a recent discussion.

On the other hand, there are indications that in quantum gravity locality is an approximate, emergent notion: the absence of fundamental local degrees of freedom is at the foundations of holography \cite{tHooft:1993dmi,Susskind:1994vu,Maldacena:1997re} and various proposals for resolving the black hole information paradox \cite{tHooft:1984kcu, Susskind:1993if,Giddings:2012gc,Bousso:2012as,Papadodimas:2012aq, Verlinde:2012cy, Maldacena:2013xja,penington2020entanglement, Almheiri:2019hni, almheiri2019entropy,Penington:2019kki,Laddha:2020kvp} rely on the existence of nonlocal quantum effects.

Understanding the fate of locality in quantum gravity is thus of primordial importance. More precisely, it remains to be understood if locality breaks down at the level of perturbation theory, or whether the aforementioned nonlocal effects are always exponentially suppressed in $1/G_N$. 

It is not straightforward to answer the question since in order to even define what we mean by locality we first need to identify candidate local observables. In a theory of gravity, these must be diffeomorphism-invariant. Defining local, diffeomorphism invariant observables in quantum gravity has proven to be challenging. This question has a rich history, see \cite{PhysRev.111.1182,Bergmann:1960wb,witten1962gravitation, Torre:1993fq,Marolf:1994wh, Khavkine:2015fwa,Marolf:1994wh,Giddings:2005id,Marolf:2015jha,Donnelly:2016rvo,Giddings:2018umg,Donnelly:2015hta} and references therein.

If the spacetime has a well-defined boundary, one approach is to define diffeomorphism invariant observables {\it relationally}, by gravitationally dressing them with respect to the boundary, but then they are not really local. Moreover in a closed universe with no boundary, this approach is not available.

An alternative would be to define observables relationally with respect to some feature of the geometry without making use of a boundary. This has been discussed in various earlier works and related ideas have been useful in the context of cosmology \cite{PhysRev.111.1182,Bergmann:1960wb,witten1962gravitation,Page:1983uc,Kuchar:1991qf,Isham:1992ms, Torre:1993fq,Marolf:1994wh,Khavkine:2015fwa,Marolf:1994wh,Giddings:2005id,Marolf:2015jha}. However, it is not clear how to give a precise mathematical definition of such observables at the quantum level, ensuring that they are exactly diffeomorphism invariant.

In this paper, we revisit the question in the framework of the AdS/CFT correspondence and we attempt to define observables dressed with respect to features of the state directly in the dual conformal field theory (CFT). An advantage of this approach is that on the boundary diffeomorphism invariance is automatically built in. A price that we pay in the construction is that the observables are defined only for a class of states.

\section{II. Algebras in time bands}

In order to investigate locality in AdS/CFT, we need to know how subregions in the bulk are encoded in the CFT. For bulk regions corresponding to the entanglement wedge of boundary subregions, this is generally understood \cite{Czech:2012bh,Almheiri:2014lwa,Jafferis:2015del}. However, for the purposes of this paper we want to find the CFT dual of a bulk subregion corresponding to a bounded causal diamond containing the candidate approximately local, diffeomorphism-invariant observable. Such regions are generally not the entanglement wedge of any boundary subregion so the mapping is of  different nature. Previous attempts to understand the CFT mapping of such regions include \cite{Balasubramanian:2013lsa,Myers:2014jia,Headrick:2014eia}. Here we will follow a different approach by focusing on the algebra of single-trace operators.

In a large $N$ holographic CFT, it is natural to define the algebra ${\cal A}$ generated by  single-trace operators in a time band ${\cal D}_{t_1,t_2}$. This was first discussed explicitly in \cite{Banerjee:2016mhh}, inspired by earlier work \cite{Papadodimas:2012aq, Papadodimas:2013wnh, Papadodimas:2013jku}. In \cite{Banerjee:2016mhh} it was proposed that the algebra ${\cal A}$ is dual to the causal wedge of the region ${\cal D}_{t_1,t_2}$ in the bulk and the commutant of ${\cal A}$ dual to the spacelike separated causal diamond in the interior. Algebras of this type have received attention recently \cite{Leutheusser:2021qhd, Leutheusser:2021frk, Witten:2021unn}.

The discussion of \cite{Banerjee:2016mhh} focused on perturbations around empty anti-de Sitter (AdS) space. In this case, the bulk geometry is homogeneous and ``featureless'' which, as we will see, introduces additional challenges in defining local diff-invariant observables. In this paper, we revisit the algebra in a time band, in cases where the bulk state is highly excited and time dependent. 

At infinite $N$, the problem can be understood in terms of QFT on a time-dependent bulk geometry, where gravitational backreaction of quantum fields can be ignored and the existence of the commutant is obvious  \footnote{If the Hamiltonian, which is an element of the time band algebra, is normalized appropriately, its commutator with bulk fields is suppressed by $1/N$.}. When considering $1/N$ corrections, the existence of the commutant is less obvious due to the gravitational Gauss law. Usually, in AdS/CFT bulk operators are gravitationally dressed with respect to the boundary, hence at order $1/N$ they do not commute with the Hamiltonian, which is an element of the algebra ${\cal A}$. This raises the question of whether the algebra ${\cal A}$ still has a commutant at subleading orders in $1/N$.

In this paper, we provide evidence for the existence of a commutant by identifying a class of operators that are gravitationally dressed  with respect to ``features of the state". As they are not dressed with respect to the boundary, these operators have vanishing commutators with the Hamiltonian, to all orders in $1/N$, thus bypassing the previous problems with the gravitational Gauss law. Here, we focus only on ensuring that bulk operators have vanishing commutators with the Hamiltonian, but an extension to all single-trace operators in ${\cal D}_{t_1,t_2}$ is necessary. We emphasize that it is really the asymptotic charges that one should be concerned with since, in the absence of gravity, bulk QFT in AdS space is manifestly local.
Understanding the algebra ${\cal A}$ in the $1/N$ expansion around empty AdS space and other static states also requires further attention \footnote{We strongly believe that our construction can be suitably generalized to include vanishing commutation with all single-trace operators, and we give a prescription to do so in \cite{long}. Concerning the acceptable class of states, the fact that the AdS vacuum is not included in this class is a feature, not a bug. Indeed, even in classical gravity, one cannot define local and diffeomorphism-invariant observables around a maximally symmetric state like the AdS vacuum.}.

The existence of a commutant for ${\cal A}$ in $1/N$ perturbation theory would imply that  information can be localized in regions of the bulk and is not visible from the boundary at the level of perturbative quantum gravity  \footnote{Other discussions of localization of information in perturbative quantum gravity, with varying conclusions, include \cite{Marolf:2008mf,Donnelly:2017jcd,Bousso:2017xyo,Donnelly:2018nbv,Jacobson:2019gnm,Giddings:2020usy,Chowdhury:2020hse, Chowdhury:2021nxw, Giddings:2021khn}.}.

\section{III. AdS/CFT setup}

We consider a holographic CFT on ${\mathbb S}^{d-1}\times$ time. The specific details of the theory are not important, but for concreteness, we can consider ${\cal N}=4$ supersymmetric Yang-Mills (SYM) theory at large $N$, large $\lambda$.

We consider a pure CFT state $\rsz$, which at large $N$ is dual to a semiclassical, time-dependent geometry.
 Various examples of such states have been discussed in the literature; see, for example,  \cite{Skenderis:2008dh,Botta-Cantcheff:2015sav,Marolf:2017kvq,Belin:2018fxe} and references therein. 
In the bulk, we may think of $\rsz$ as a time-dependent coherent state. The state may eventually collapse into a black hole, though this is not central to the discussion. We can also consider black hole microstates with semiclassical time dependence in the region behind the horizon \cite{Kourkoulou:2017zaj, Almheiri:2018ijj,Cooper:2018cmb,Miyaji:2021ktr}.

Any such state can be expanded in the basis of CFT energy eigenstates as
\be
\rsz =\sum_i c_i |E_i\rangle
\ee
We consider states with $\lsz H \rsz \sim O(N^2)$. Given the nontrivial time dependence of the bulk geometry, such states will have energy variance
\be
\label{vardef}
(\Delta H)^2 = \lsz H^2 \rsz - \lsz H\rsz^2
\ee
also of $O(N^2)$  \footnote{This can be seen from the inequality $\frac{1}{2}|\braket{ [H,A]}|\leq \Delta H \cdot\Delta A$, since large $N$ factorization implies $\frac{\Delta A}{ \braket{ A}}\sim O(N^{-1})$. Notice that if the variance is parametrically larger than $O(N^2)$, the state may not have a good semiclassical interpretation.}. 

We want to construct approximately local diffeomorphism-invariant observables on the geometry dual to this state. A standard approach is based on the Hamilton-Kabat-Lifschytz-Lowe (HKLL) construction \cite{Banks:1998dd, Bena:1999jv,hamilton2006local, hamilton2006holographic, hamilton2007local, hamilton2008local,Heemskerk:2012mn}, which expresses the desired bulk operator in terms of smeared CFT local single-trace operators. For example, for a scalar field and at large $N$ we get an expression of the form
\be
\label{hkll}
\hkll(t,r,\Omega) = \int_{\rm bdry} dt' \,d\Omega'_{d-1} K(t,r,\Omega; t',\Omega') {\cal O}(t',\Omega')
\ee
where $K$ is related to a Green's function of the Klein-Gordon operator on the bulk geometry. See \cite{hamilton2006holographic} for explicit expressions for the kernel $K$ in the AdS vacuum. Notice that in order to define the kernel $K$ we have to choose a coordinate system in the bulk, for example, using Fefferman-Graham or harmonic gauge, which is defined by making use of the asymptotic boundary.

The operator \eqref{hkll} is defined in the CFT, hence obviously invariant with respect to bulk diffeomorphisms. To leading order at large $N$, \eqref{hkll} behaves like a local operator in the bulk, i.e. it commutes with other operators at bulk spacelike separation.

At subleading order in $1/N$, such commutators generally become nonzero. In order to preserve the local behavior of the reconstructed bulk operator, the expression \eqref{hkll} needs to be corrected order by order in $1/N$ by adding to \eqref{hkll} other single- and multitrace contributions \cite{Kabat:2011rz,Kabat:2012av,Heemskerk:2012mn}.

However, there is a universal nonvanishing $1/N$ commutator that cannot be corrected this way, in particular, the  commutator of \eqref{hkll} with the CFT Hamiltonian. The CFT Hamiltonian is dual to the Arnowitt-Deser-Misner Hamiltonian in the bulk, which can be defined in a spacelike separated region relative to the bulk point where \eqref{hkll} is localized. 

The physical origin of this effect is the gravitational Gauss law: acting with \eqref{hkll} will generally create or destroy a particle in the bulk, thus changing the energy of the state, which can be immediately measured at spacelike infinity by $H$. Another way to think about it is that the operator \eqref{hkll} is defined relationally with respect to the boundary: the coordinate system used to compute $K$ in \eqref{hkll} is defined by some gauge fixing condition that makes use of the asymptotic boundary. Time translations by the CFT Hamiltonian then also time translate operator \eqref{hkll}. We can also think in terms of (smeared) gravitational Wilson lines connecting the bulk operator to the boundary, which make it diffeomorphism invariant at the price of making it nonlocal \cite{Anand:2017dav,Castro:2018srf,Chen:2019hdv,Giddings:2019hjc}. The commutator with $H$ is nonzero as $H$ picks up the Wilson line.

Our goal is to improve the locality properties of \eqref{hkll} by moving the gravitational dressing from the boundary to the state. From a technical point of view, we will find a CFT operator $\drop$ that obeys two properties: (i) $[H,\drop]=0$ to all orders in $1/N$ and (ii) to leading order at large $N$, correlators of $\drop$ agree with those of $\Phi$. The latter condition  guarantees that the operator acts in a desirable way. We will comment on commutators with other  asymptotic charges below.

\section{IV. Time-shifted states and return amplitude}

Starting with the state $\rsz$ we consider the one-parameter family of states
\be
|\Psi_T\rangle = e^{-i T H} \rsz \qquad T\in {\mathbb R}.
\ee
In the bulk, the states $|\Psi_T\rangle$ are related to $\rsz$ by a large diffeomorphism. They are {\it different} quantum states, even though they are related by symmetry. From the point of view of the phase space of gravity in AdS space, they correspond to different phase space points.

As discussed earlier, in the bulk we can think of $|\Psi_T\rangle$ as coherent states. Based on general intuition about the overlap of coherent states, we expect an overlap of the form
$\lsz \Psi_T\rangle = e^{-{1\over \hbar} f(T)}$. 
In AdS/CFT the effective $\hbar$ is proportional to $1/N^2$, hence we expect
\be
\label{decaybulk}
\lsz\Psi_T\rangle = e^{-N^2 f(T)}
\ee
It is not straightforward to compute $f(T)$ from semiclassical gravity; see \cite{Papadodimas:2015jra} for a discussion on nearby states. In principle, $f(T)$ can be computed by using a Euclidean preparation of the states \cite{Belin:2018fxe}. The computation of $f(T)$ directly from Lorentzian bulk data is an interesting challenge.
Microscopically we have
\be
\label{decaybdry}
\lsz\Psi_T\rangle = \sum_i |c_i|^2 e^{-i T E_i }
\ee
and the suppression \eqref{decaybulk} comes from the summation over a large number of phases.

If the bulk state has no periodicities, we expect $f(T)$ to increase as we increase $T$. On the other hand, an estimate of \eqref{decaybdry} shows that the decay will saturate at some point. Indeed, the nontrivial overlaps \eqref{decaybdry}  means that it is not correct to think that of all states $|\Psi_T\rangle$ simultaneously as being independent, see also \cite{Papadodimas:2015xma, Papadodimas:2015jra,Chakravarty:2020wdm} for related discussions. One aspect of this can be understood in terms of Poincar\'e recurrences that will happen at very large $T\sim \mathcal{O}(e^{e^{N^2}})$. In this paper, we will be interested in much earlier timescales so it will be sufficient to treat the states as quasiorthogonal since all overlaps will be exponentially small.

Starting with the state $\rsz$ we define the  code subspace
\be
\label{code0}
{\cal H}_{ 0} = {\rm span}\{|\Psi_0\rangle,{\cal O}(t,\Omega) |\Psi_0\rangle,...,{\cal O}_1(t_1,\Omega_1)...{\cal O}_n(t_n,\Omega_n)\rsz\}
\ee
generated by acting on $\rsz$ with a small number ($n\ll N$) of single-trace operators \footnote{More precisely, we would have to give a small smearing to the single-trace operators in order to avoid UV divergences.}. We also define the projector $P_0$ on this subspace.

A similar code subspace can be defined for each of the time-shifted states
\be
\label{codet}
{\cal H}_{T}= {\rm span}\{|\Psi_0\rangle,{\cal O}(t,\Omega) |\Psi_T\rangle,...,{\cal O}_1(t_1,\Omega_1)...{\cal O}_n(t_n,\Omega_n)|\Psi_T\rangle\}
\ee
with the corresponding projector $P_T$. We have
\begin{equation}
 P_T = e^{-i TH } P_0 e^{i TH}   
\end{equation}
and in particular, it is important to keep in mind that  $P_T \neq P_0$.

\subsection{The return amplitude}

We now examine the $T$ dependence of the overlap \eqref{decaybdry} in more detail. Consider the quantity
\be
\label{radef}
R(T) = |\lsz e^{-i T H} \rsz|^2
\ee
called the return amplitude. It is closely related to the spectral form factor \footnote{The spectral form factor coincides with the return amplitude for the thermofield double (TFD) state $\ket{{\rm TFD}}$ with $H=H_L+H_R$.}, which has been extensively discussed recently in the context of the black hole information paradox; see, for example, \cite{Cotler:2016fpe}.

In general, it is difficult to compute \eqref{radef}.  As discussed earlier, in principle, we should be able to capture the early time, large $N$ behavior of \eqref{radef} in terms of overlaps of time-shifted coherent states. We present some more detailed computations in \cite{long}. Here we notice that for very early times
\be
\label{earlyr}
R(T) = e^{-(\Delta H)^2 T^2}.
\ee
For states with an energy variance of $O(N^2)$, this is a very fast decay of the order
\be
\label{decay}
R(T)=e^{-\alpha T^2 N^2},
\ee
where the constant $\alpha$ is $O(N^0)$ and depends on the specific state we are considering \footnote{This decay prevails up to timescales of order $T\sim \tilde{\alpha} N^0$, but may only be valid for for $\tilde{\alpha}\ll 1$. At times scales $T\sim \mathcal{O}(1)$, the return amplitude will behave similarly to \eqref{decaybulk} with a function $f(T)>0$.}. The decay \eqref{decay} of $R(T)$ is parametrically faster than thermalization, whose timescale  is typically of order  $O(N^0)$. 
For a system with no degeneracies,  \footnote{Systems like ${\cal N}=4$ SYM will have degeneracies due to symmetries, however, the number of degenerate states is exponentially smaller than the number of all states in the high-energy  sector of the theory.}
\be
\label{plat}
\overline{R} = \lim_{t_*\rightarrow \infty}{1\over 2t_*} \int_{-t_*}^{t_*} \,dT\,R(T) = \sum_i |c_i|^4.
\ee
For the type of states we are considering, the rhs is exponentially small, scaling as $e^{-\alpha' N^2}$, where $\alpha'$ is an $O(1)$ constant which depends on $\rsz$.

Between the initial decay \eqref{decay} and the long-time plateau \eqref{plat}, there may be other interesting intermediate regimes, which have received attention in connection to quantum chaos \cite{Shenker:2013pqa,Saad:2018bqo}. What is necessary in the following discussion is  that already from timescales of order $t\sim N^0$ and generally, the return amplitude remains exponentially suppressed in $N^2$ for a long time.

Here we notice that the return amplitude obeys the obvious property
\be
\label{timetrans}
\langle \Psi_{t_0} |\Psi_{t_0+T}\rangle = \langle \Psi_0|\Psi_T\rangle
\ee
This means that even if the bulk geometry appears to be static at the semiclassical level, the return amplitude can still decay like \eqref{decay} if the state had a period of manifest bulk time dependence in the far past. This observation is relevant, for example, in the case of a black hole formed by gravitational collapse.

The exponential decay \eqref{decay} can be extended to more general correlators of the form $\lsz {\cal O}(t_1)\ldots {\cal O}(t_n) |\Psi_T\rangle$, where ${\cal O}$ are single-trace operators. We expect
\begin{equation}
\label{cordecay}
    \begin{split}
 |\lsz {\cal O}(t_1)\ldots {\cal O}(t_n) |\Psi_T\rangle| &= 
        \\
    |\lsz {\cal O}(t_1)\ldots {\cal O}&(t_n) \rsz| e^{-\alpha T^2 N^2}
  +  [~{\rm subleading}~]     
    \end{split}
\end{equation}
The meaning of ``subleading" is as follows: we will need to insert the expression above inside an integral over $T$, which will be computed by a saddle point method as $N\rightarrow \infty$. The claim is that the subleading terms above contribute only at $O(1/N)$ to that integral. The intuition is that since in the large $N$ limit $\rsz$ and $|\Psi_T\rangle$ are different semiclassical states, they cannot be connected by the action of a small number of single-trace operators. Further evidence for the behavior \eqref{cordecay} in various examples will be given in \cite{long}.

\begin{figure}[h]
    \centering
    \includegraphics[width=0.45 \textwidth]{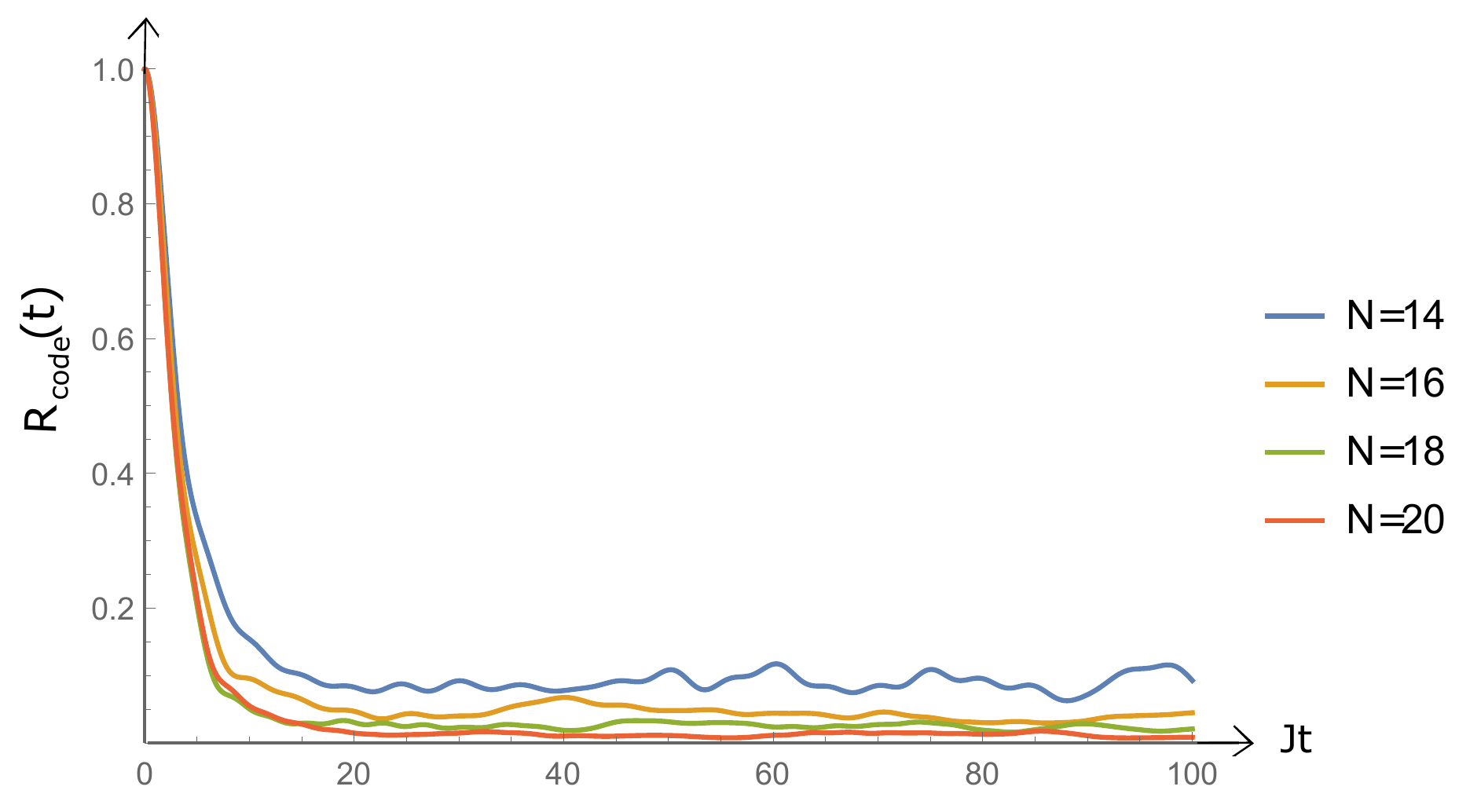} 
\caption{A numerical study of the decay of $R_{code}(t)$ as a function of $t$ in the SYK model. The state $\rsz$ is one of the Kourkoulou-Maldacena microstates \cite{Kourkoulou:2017zaj}. The code subspace is obtained by acting on it with a set of fermionic operators. In the plot $d_{\rm code}=8$. More details will be provided in \cite{long}.}
 \label{decayofcode}
    \end{figure}

Another way of looking at this is that any state in the code subspace \eqref{code0} has an exponentially small overlap with any state in the code subspace \eqref{codet}. This can be captured by \footnote{Various useful inequalities can be derived: for a Hermitian operator ${\cal O}$ with eigenvalues $\lambda_i$, and if $[P_0,{\cal O}]=0$, we have
 $ |\lsz {\cal O} \ket{\Psi_T}|^2\leq \sqrt{\Tr[{\cal O}^4]} \sqrt{\Tr[P_T P_0]}$ and $|\lsz {\cal O} \ket{\Psi_T}|^2\leq {\rm max} (\lambda_i^2)\, \Tr[P_T P_0]$ }
\be
\label{orthcode}
R_{code}(T) = \frac{1}{d_{code}} Tr[P_T P_0]
%={\Tr[P_T P_0] \over d_{code}} 
= O(e^{-\alpha T^2 N^2 })
\ee
for the relevant timescales. Here $d_{code}$ is the dimensionality of the code subspace.
We provide some numerical evidence for this in the case of the Sachdev-Ye-Kitaev (SYK) model in Fig. \ref{decayofcode}.

\section{V. The operators}

We now introduce \footnote{Operators of this type were discussed in a related context and applied to the eternal black hole in \cite{Papadodimas:2015xma}.} operators $\hat{\Phi}$ with the desired properties \footnote{Since $[\Phi,P_0]=0$ in the code subspace, we could have defined operators with the same action on the code subspace as \eqref{defop}, using a single projector on the left of $\Phi$. However, in that case, the operators would have extraneous matrix elements associated with orthogonal subspaces to ${\cal H}_0$.}
\be
\label{defop}
\drop= c\int_{-t_*}^{t_*}dT\,\, e^{-i T H } P_0 \Phi P_0 e^{i T H }
\ee
Here $\Phi$ is a usual boundary-dressed operator like \eqref{hkll}, $P_0$ is the projector on the code subspace \eqref{code0}, $t_*$ is a timescale that needs to be at least $O(N^0)$, and $c$ is an overall normalization constant
\be
\label{normalization}
c^{-1} = \int_{-t_*}^{t_*} dT \lsz P_T \rsz
\ee
We now prove the two desired properties.

\subsection{A. Vanishing commutator with $H$ to all orders in 1/N}

We start with
$$ [H,\drop] =-i{d\over ds}\big(e^{i s H } \drop e^{-is  H }\big) \Big|_{s=0}.$$
From \eqref{defop} and by a change of variables we obtain 
$$
[H,\drop]=-i{d\over ds}\big(c\int_{-t_*-s}^{t_*-s}dT\,\, e^{-i T H } P_0 \Phi P_0 e^{i T H }\big)\Big|_{s=0}
$$
which reduces to boundary terms
$$
[H,\drop] = i c (P_{t_*} \Phi(t^*) P_{t_*} - P_{-t_*} \Phi(t^*)  P_{-t_*} ).
$$
If we select $t_*$ to be large enough, then by using \eqref{orthcode} we find that when inserted in a correlator inside the code subspace of the state $\rsz$ we arrive at
\be
\label{hcommut}
[H,\drop] = O(e^{-\gamma N^2}),
\ee
where $\gamma$ is positive and $O(N^0)$. This proves the first desired property.

\subsection{B. Similar action as HKLL operators}

We also want to make sure that $\hat{\Phi}$ has the same correlators as the HKLL operator \eqref{hkll} to leading order at large $N$. For that we consider
\begin{equation} \label{eq20}
    \begin{split}
\lsz&{\cal O} ...\drop ... {\cal O} \rsz 
        \\
= &c \int_{-t_*}^{t_*} dT\, \lsz {\cal O}...e^{-iT H } P_0 \Phi P_0 e^{i TH } ... {\cal O} \rsz
        \\
        = &c \int_{-t_*}^{t_*} dT\, \lsz {\cal O}...P_T (e^{-iTH } \Phi e^{i TH })P_T ... {\cal O} \rsz
        \\
       =&c \int_{-t_*}^{t_*} dT\, \lsz {\cal O}...P_0 P_T (e^{-i TH } \Phi e^{iT H }) P_T P_0... {\cal O} \rsz .
    \end{split}
\end{equation}
In the last line, we insert the projectors $ P_0$ as we are free to do so.
Now from \eqref{orthcode} we see that the integrand will be exponentially suppressed as $|T|$ increases. In the large $N$ limit, we can estimate the integral by a saddle point method, which is dominated by the $T=0$ contribution \footnote{Note that the correlator appearing in \eqref{eq20} can be written as a product of a correlator \eqref{cordecay} and a conjugated correlator, such that the phase factor cancels and gives a time dependence following \eqref{orthcode}.}. Using \eqref{cordecay} and \eqref{normalization} we find that 
\be
\lsz {\cal O}...\drop ... {\cal O} \rsz =
\lsz {\cal O}...\Phi ... {\cal O} \rsz  + O(1/N),
\ee
as desired.

Notice that if we apply the operator $\drop$ to one of the time-shifted states, then as long as $|T| < t_*$, we find
\begin{multline}\label{shiftedaction}
  \langle \Psi_T |{\cal O}...\drop ... {\cal O} |\Psi_T\rangle= 
  \\
   \langle \Psi_T| {\cal O}...(e^{-iT H } \Phi e^{i TH }) ... {\cal O} |\Psi_T\rangle  + O(1/N)
\end{multline}
which we will discuss below. To make \eqref{shiftedaction} more manifest, we can also write $\drop$ as
\begin{equation}
  \drop = c \int_{-t_*}^{t_*} dT~ P_T (e^{-i T H} \Phi e^{i T H })P_T . 
\end{equation}

\subsection{C. Interpretation and comments}

To leading order at large $N$ the operator \eqref{defop} acts like the HKLL operator \eqref{hkll}. However, its commutator with $H$ is zero to all orders in $1/N$.  The existence of these operators provides evidence that the algebra of single-trace operators in a short time band can have a nontrivial commutant when acting on heavy, time-dependent states. 
The vanishing of the commutator with $H$  happens because \eqref{defop} is gravitationally dressed, not with respect to the boundary, but instead with respect to the time dependence of the state. Suppose, for example, that in the state $\rsz$ we have a supernova explosion taking place at $t=0$ and the operator \eqref{hkll} is selected so that it acts near the explosion. In the state $|\Psi_T\rangle$ the explosion will take place at $t=-T$. From Eq. \eqref{shiftedaction} we see the operator $\drop$ will again act in the bulk near the new location of the supernova explosion. Hence, one and the same operator $\drop$ knows how to always act at the moment of the explosion for the entire family of states $|\Psi_T\rangle$, $|T|< t_*$.

It is not possible to apply the same logic in empty AdS space or other static states,  as there are no time-dependent features in the bulk to be used as a clock to define a moment in time where the operator acts. Technically, the return amplitude for such states does not exhibit the rapid decay \eqref{decaybulk}.

More generally, we need to make \eqref{hkll} commute with all boundary symmetry generators, besides $H$. In case there is only conformal symmetry, we should consider a generalization of the form
\be
\drop = c\int_{B} d\mu(g) U(g)P_0 \Phi P_0 U(g)^{-1},
\ee
where $d\mu(g)$ is the Haar measure on $SO(d,2)$ and $B$ is a reasonably sized connected submanifold of $O(d,2)$ containing the identity. 
The commutator with conformal generators will then be given by operators in the code subspace of states $U(g_*)|\Psi_0\rangle$, where $g_*$ lies on the boundary $\partial B$. For the construction to work in this generalization we must make sure that the overlaps
\beq
\label{genoverlap}
R(g) = |\langle \Psi_0|U(g)|\Psi_0 \rangle|^2 
\eeq
decay exponentially in the geodesic distance of $g$ from the identity. We expect this to be true for states that break all symmetries at the semiclassical level \footnote{For compact symmetries, such as rotations, $R(g)$ will have recurrences every $2\pi$. Hence along the compact directions we take $g_* \sim O(1)< 2\pi$.}. The expression given as $R(g)$ is an interesting generalization of the return amplitude \eqref{radef} that would be interesting to study further.

Suppose that the bulk state can be thought of as being made out of two distant, weakly interacting  subsystems. As an idealization, we can model it by two noninteracting CFTs with total Hamiltonian $H=H_L+H_R$. The full system is in a pure state $\rsz$ which may be entangled, but we assume the pattern of entanglement is generic. We consider the two-parameter family of time-shifted states 
$$e^{-i (T_L H_L + T_R H_R)} \rsz.$$
We start with an HKLL operator $\Phi$ on the left system, which will commute with $H_R$ but not $H_L$. Then we can consider the following generalization of the operators \eqref{defop}
\be
\drop = c\int dT_L dT_R e^{-i ( T_L H_L + T_R H_R )} P_0 \hkll  P_0 e^{i (T_L H_L  + T_R H_R)}
\ee
using $P_0 = P_0^L \otimes P_0^R$ and  $[\hkll, P_0^R]=0$, then
\be
\drop = c\int dT_L  e^{-i T_L H_L} P_0^L \hkll P_0^L  e^{i T_L H_L}\otimes \int  dT_R P_{T_{R}}^R
\ee
The resulting operator commutes with both $H_L$ and $H_R$ on the relevant space of states. In this case, instead of saying that the operator is dressed with respect to the overall time dependence of the entire system, we can say that it is actually dressed with respect to the time dependence of the ``left'' subsystem \footnote{In states with special entanglement the generalized return amplitude $|\lsz e^{-i (T_L H_L + T_R H_R )} \rsz|^2$ may not decay in both $T_L$ and $T_R$. For example, in the TFD state it is constant along the line $T_L=-T_R$. In those cases we cannot set both commutators with $H_L,H_R$ to zero.}.

\section{VI. Black Hole microstates}

Suppose $\rsz$ is a black hole microstate with energy variance of $O(N^2)$. Such states are microscopically time dependent.
For some of these states, the time dependence may be visible at the semiclassical level, for example, in states with end of the world branes behind the horizon \cite{Kourkoulou:2017zaj, Almheiri:2018ijj,Cooper:2018cmb,Miyaji:2021ktr}. In those cases we can say that operators \eqref{defop} are gravitationally dressed with respect to the end of the world brane. For more general microstates with energy variance $O(N^2)$ it may not be easy to understand the time dependence semiclassically in the bulk. Notice, however, that the mathematical properties of \eqref{defop} only depend on the rapid decay of the return amplitude, which is expected to be true even in those states. Hence we can say that even in those states operators \eqref{defop} are dressed with respect to the overall time dependence of the state. 

Consider a model of black hole evaporation where a black hole in a holographic CFT, slowly evaporates into Hawking radiation absorbed by a nongravitational QFT. After the Page time, it is believed that part of the black hole interior is encoded in the radiation. Suppose we start with an operator $\Phi$ in the black hole interior. We assume that $\Phi$ is gravitationally dressed with respect to the CFT, hence it does not commute with $H_{\rm CFT}$. In this case the two systems are weakly interacting but highly entangled.  A protocol like the one described in the previous subsection allows us to promote $\Phi$ into an operator $\hat{\Phi}$ which acts similarly on the code subspace of the state but has vanishing commutators with $H_{\rm CFT}$ to all orders in $1/N$. The operator $\hat{\Phi}$ is  gravitationally dressed with respect to the radiation \footnote{For the {\it exact} operator in the island the commutator with $H_{\rm CFT}$ may be exactly zero, here we simply point out that there is no contradiction with perturbative diffeomorphism invariance.}. This suggests that there is no inconsistency between the gravitational Gauss law and the island prescription, and may be useful \cite{long} in resolving paradoxes raised in \cite{Geng:2021hlu}. 

\section{VII.Discussion}

In this paper, we have presented a construction of CFT operators that act as local bulk operators in a code subspace, but commute with the Hamiltonian to all orders in the $1/N$ expansion. The gravitational interpretation of such operators is that they are bulk local operators that are gravitationally dressed to features of the state, in particular its time dependence. Because the operators are constructed directly in the CFT, they are manifestly diffeomorphism invariant. We conclude with some open questions.

It would be interesting to understand if there is a natural way to identify operators whose commutators is zero to all orders in $1/N$ with both $H$ and other single-trace operators in the time band, thus proving the conjecture that the time band algebra has a commutant in the $1/N$ expansion.

It would also be interesting to understand how to analyze states with very small energy variance, for instance typical black hole microstates in the sharp microcanonical ensemble \cite{Papadodimas:2017qit,deBoer:2018ibj, DeBoer:2019yoe}, energy eigenstates, or even empty AdS space. In these cases the return amplitude does not decay fast enough and the construction \eqref{defop} cannot be applied. These are also the states where there is no semiclassical feature of the state to dress with respect to, or in other words there is no bulk observer. It may be interesting to clarify the role of the observer, perhaps as in \cite{Chandrasekaran:2022cip}, toward identifying a commutant for the time band algebra in those states.

\vspace{0.3cm}

\section{Acknowledgments}

We would like to thank S. Banerjee, J. de Boer, P. Caputa, E. Kiritsis, M. Mirbabayi, A. Parnachev, S. Raju, E. Verlinde, G. Vos, S. Wadia and S. Zhiboedov for useful discussions. The work of A. B. is supported by the NCCR 51NF40-141869 The Mathematics of Physics (SwissMAP). E.B. and N.V. would like to thank CERN-TH for hospitality during the preparation of this work.

\bibliography{references.bib} 

\end{document}